\documentclass[aps,pre,showpacs,twocolumn,superscriptaddress,longbibliography]{revtex4-1}
\usepackage[pdftex]{graphicx}
\usepackage[pdftex,bookmarks,colorlinks,breaklinks]{hyperref} 
\usepackage{amsmath,amssymb}
\hypersetup{linkcolor=red,citecolor=blue,filecolor=blue,urlcolor=blue} 
\hypersetup{pdftoolbar=true,pdfpagemode=UseNone,pdfstartview=FitH,colorlinks=true,extension=}

\newcommand{\CT}{\ensuremath{{\cal{CT}}}}
\newcommand{\calT}{\ensuremath{{\cal{T}}}}
\newcommand{\calC}{\ensuremath{{\cal{C}}}}
\newcommand{\calH}{\ensuremath{{\cal{H}}}}
\newcommand{\evec}{\vec{\text{e}}}
\newcommand{\gammanb}{\gamma_{\text{nb}}}

\begin{document}
\title{Microwave limiters implemented by coupled dielectric resonators\\
	based on a topological defect mode and \CT-symmetry breaking}
\author{Mattis Reisner}
\affiliation{Universit\'{e} C\^{o}te d'Azur, CNRS, Institut de Physique de Nice (INPHYNI), 06108 Nice, France, EU}
\author{Fabrice Mortessagne}
\affiliation{Universit\'{e} C\^{o}te d'Azur, CNRS, Institut de Physique de Nice (INPHYNI), 06108 Nice, France, EU}
\author{Eleana Makri}
\affiliation{Wave Transport in Complex Systems Lab, Department of Physics, Wesleyan University, Middletown CT-06459, USA}
\affiliation{Department of Physics, Queens College of the City University of New York, Flushing NY-11367, USA}
\author{Tsampikos Kottos}
\affiliation{Wave Transport in Complex Systems Lab, Department of Physics, Wesleyan University, Middletown CT-06459, USA}
\author{Ulrich Kuhl}
\affiliation{Universit\'{e} C\^{o}te d'Azur, CNRS, Institut de Physique de Nice (INPHYNI), 06108 Nice, France, EU}


\begin{abstract}
	We present a microwave realization of a reflective topological limiter based on an explicit self-induced violation of a charge-conjugation (\CT) symmetry. 
	The starting point is a bipartite structure created by coupled dielectric resonators with a topological defect placed at the center and two lossy resonators placed on the neighboring sites of the defect. 
	This defect supports a resonant mode if the \CT-symmetry is present, while it is suppressed once the symmetry is violated due to permittivity changes of the defect resonator associated with high irradiances of the incident radiation.
	This destruction leads to a suppression of transmittance and a subsequent increase of the reflectance while the absorption is also suppressed. 
\end{abstract}

\pacs{05.45.Mt}

\maketitle

\section{Introduction}
Limiters are an important ingredient of modern technologies, where communication via electromagnetic waves requires compact devices which, in turn, lead to a more complex electromagnetic environment. 
Testaments of such complexity are developments in Wi-Fi, Bluetooth, mobiles as well as near field communications (NFC) spanning frequency ranges from MHz (radiowaves) to several tens of GHz (microwaves). 
With the introduction of the Internet of Things (IoT) and 5G the need for next generation limiters will become even more pronounced \cite{hue19}. 
Limiters should protect the communication devices from detrimental high incident powers or fluence. 
There are different types of limiters ranging from diode limiters \cite{bil97}, gas-tube limiters based on generating plasmas \cite{cro13,cpi12} or limiters based on superconducting technologies \cite{boo03,lin99b}, each with their advantages and disadvantages. 
Recently the idea of using topological structures for the realization of a new generation of resilient limiters has been proposed \cite{mak14,mak15} and already experimentally studied in the microwave regime \cite{kuh17}. 
We will present here a microwave realization of a reflective limiter based on the \CT-symmetry breaking in a bipartite structure which has been proposed in Ref.~\onlinecite{mak18}.
As opposed to previous proposals, the topological reflective limiter is robust against fabricational errors, while at the same time is resilient to high power incident radiation that might lead to a destruction of existing limiting schemes.

The structure of this paper is as follows. 
In the next section we present the experimental setup including details on the dielectric resonators, their choices and preparation. 
Section III presents the experimental results on the limiter itself, especially its transmittance, reflectance, and absorption properties.

\section{Experimental setup and \CT-breaking}

The experiment is based on evanescently coupled dielectric resonators sandwiched between two metallic plates. 
For details on the coupling mechanism and its description we refer to Refs~\onlinecite{bar13a,bel13b}. 
These resonators have already been used in linear chain setups to demonstrate topological defects \cite{pol15,kuh17}, Dirac oscillators \cite{fra13} or chiral ensembles \cite{arXreh19} as well as in two dimensional structures like graphene \cite{bar13a,bel13b,bel13a} or Lieb lattices \cite{pol17}. 
Of course in actual limiters one has to incorporate non-linear mechanisms that allow for a distinction between low and high incident power -- the latter turning the limiter to its ``off'' state --, but for a proof of principle experiment it is sufficient to perform the permittivity variation parametrically, as carried out here.
Non-linearities can be implemented in this microwave setup \cite{arXrei19} and can be treated theoretically and numerically \cite{hen99}.

\begin{figure}
	\centering 
	\includegraphics[width =\linewidth]{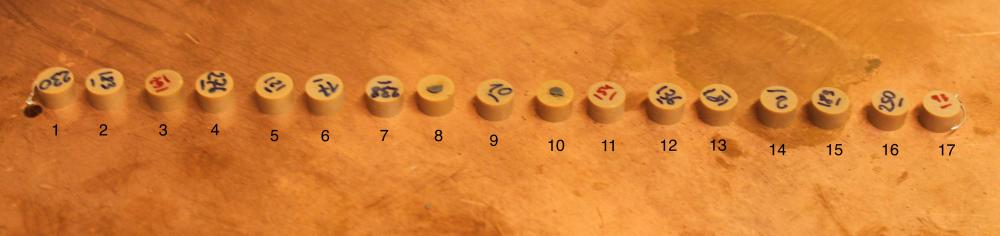}
	\caption{\label{fig:photochain}
		The experimental setup showing a 4-D-4 dimer chain with curled kink antennas coupled to the utmost left and right resonators. 
		The neighboring resonators of the topological defect ($n_d=$9) have the additional absorbing elastomers on top of them.
	}
\end{figure}

Figure~\ref{fig:photochain} shows a photograph of the realized topological limiter using in total 8 dimers with intra-dimer distances of $d_1=11$\,mm (coupling $t_1=41.4$\,MHz) and inter-dimer distances $d_2=13$\,mm (coupling $t_2=21.2$ MHz) with an average eigenfrequency of the resonators of $\nu_0=6.069$\,GHz (see also appendix~\ref{sec:singlereso}). 
At the center of the chain a topological defect is placed by repeating an inter-dimer distance $d_2$ generating a 4-D-4 structure. 
Since the central resonator, and therefore the defect is between two topologically distinct regions, it results in a topologically protected bound state \cite{sch13} which occurs exactly at $\nu_0$, the eigenfrequency of the defect resonator, in the spectra. 
This mode, also called zero-mode is localized at the defect ($n_d$).
As the eigenfrequency of the defect resonator will be varied parametrically later on we denote it by $\nu_d$. 
Note, that in actual implementations of the set-up of fig.~\ref{fig:photochain}, the intensity of the incident radiation will be responsible for the variations in $\nu_d$. 
For example, in the optical framework, the value of $\nu_d $ will change due to a Kerr non-linearity. 
In the microwave domain we can change $\nu_d$ by inductively coupling the defect resonator with a non-linear circuit.
On the leftmost and rightmost resonators a curled kinked antenna is placed to guarantee strong coupling to the system (see also fig.~\ref{fig:photo_ant}). 
Finally, we have incorporated large Ohmic losses $\gammanb$ to the two resonators on the left and right of the defect. 
The loss has been incorporated by placing strong absorbing material on top of the surface of these resonators (see also fig.~\ref{fig:photo_abs}).

\begin{figure}
	\centering
	\includegraphics[width = \linewidth]{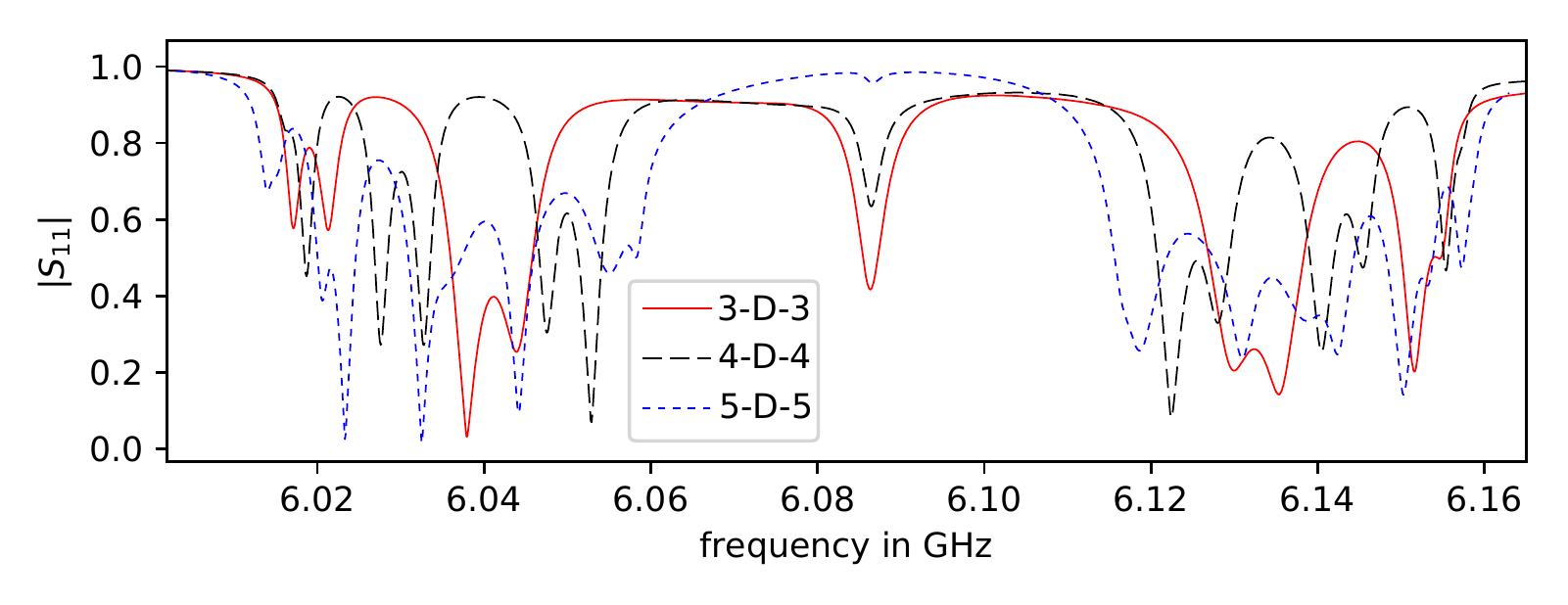}
	\caption{\label{fig:different_N}
		Measured reflection $|S_{11}|$ with the strongly coupled antennas for defect dimer chains with different total numbers $N$ of dimers and a defect in the center ($N$-D-$N$: $N$-dimers-Defect-$N$-dimers).}	
\end{figure}

The infinite dimer structure generates a band-gap given by $ \nu_0-|t_1 - t_2|<\nu < \nu_0+ |t_1 - t_2|$. 
For the chosen setup this corresponds to $ 6.049$\,GHz $<\nu< 6.089$\,GHz. 
The band structures can be seen in fig.~\ref{fig:different_N} where by increasing the number of dimers $N$ at each side of the defect the band gap is approaching $ 6.060$\,GHz $<\nu< 6.115$\,GHz. 
The shift of the gap is due to the fact that at very small distances the presence of the second resonator gives rise to an additional shift of the bare frequency \cite{arXreh19}. 
This effect has not been taken into account in the theoretical description. 
While the gap becomes better and better defined for increasing number of dimers the defect state gets less and less pronounced due to the increase of the time-delay leading to an enhancement of the absorbed energy in the defect resonator, as even without any additional absorbing material the resonator has a non-negligible absorption ($\gamma=1.8$\,MHz). 
The $N=4$ value seems to be a reasonable compromise to have a well established defect resonance and a well defined band gap. 
Thus we will concentrate on the 4-D-4 dimer setup for the limiter. 

\begin{figure}
	\centering 
	\includegraphics[width =\linewidth]{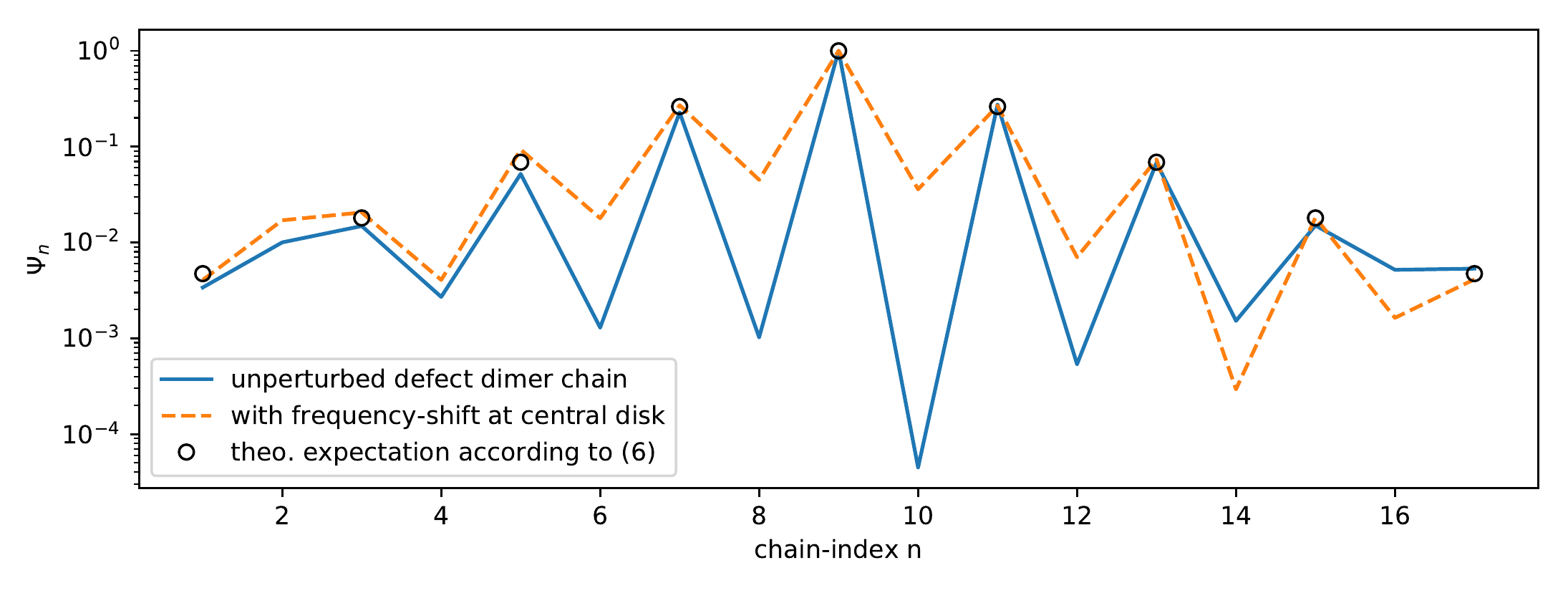}
	\caption{\label{fig:eigenstate}
		Blue solid line: Measured state wave function $\Psi$ for the unperturbed chain [$\nu_d = 6.069$\,GHz, see eq.~(\ref{eq:wavefunction})]
		Dashed orange line: Chain with a frequency-shift at the central resonator ($\nu_d = 6.092$\,GHz), 
		Open circles: Eigenfunction for infinite chain. 
		All are normalized to $|\Psi(n_d)|S=1$ at the defect.
	}
\end{figure}

In case of no absorption, the system is time-reversal symmetric, thus the Hamiltonian $H$ commutes with the time-reversal operator $\calT$, $\calH\calT=\calT\calH$, where $\calT^2=1$. 
Furthermore, as the lattice has a bipirtite structure, i.e.~the couplings alternate between small and large values, and next-nearest neighbor couplings are negligable, the effective Hamiltonian of the system commutes with the chiral operator \cite{arXreh19}.
This can be easily seen if one writes the Hamiltonian in the position basis writing first the odd resonators and then the even ones.
The Hamiltonian for such a situation may be written as
\begin{equation}\label{eq:chiham}
\calH=\nu_0\cdot \mathbf{1}_M+\left(\begin{array}{cc}
0 & A \\
A^\dag & 0
\end{array}\right)\,,
\end{equation}
where the diagonal blocks belong to the two subsystems, and the off-diagonal blocks describe the interaction, $M$ is the number of resonators and $\mathbf{1}_M$ is the $M$ dimensional diagonal matrix. 
From the above discussion it becomes evident that the chiral symmetry is protected when all resonators have the same eigenfrequency $\nu_0$.
Thus it exist an chiral operator $\calC$ which is {\em anti}commuting with $\calH$, $\calH\calC=-\calC\calH$, with $\calC^2=1$ \cite{haa18,zir96,alt97c,bee15}.
In total the Hamiltonian of this system possesses an anti-unitary charge-conjugation symmetry \cite{su79,sch13,mal15}
\begin{equation}\label{eq:hamiltonian}
\left\lbrace \mathcal{CT} ,\mathcal{H} \right\rbrace \equiv \mathcal{CTH} + \mathcal{HCT} = 0 ,
\end{equation}
This results in pairs of eigenfrequencies $\nu_0 + \nu_m$, $\nu_0 - \nu_m^{*}$. 
Due to this symmetry the unpaired eigenmode at $\nu_0$ is stabilized and in the limit of $N \rightarrow \infty$ its wavefunction is exponentially localized and given by \cite{sch13,pol15}
\begin{equation}\label{eq:wavefunction}
\Psi_n \sim
\begin{cases}
\frac{1}{\sqrt{\xi}}\text{e}^{-\frac{|n - n_d|}{\xi}} & \text{for } n \text{ odd}, \\
0 & \text{for } n \text{ even}, \\
\end{cases}
\end{equation}
where $ \xi = 1/ \ln(t_1/t_2)$ is the so-called localization length of the mode \cite{has10}.
If we include now a global absorption, i.e., the same $\gamma$ for all resonators the mode will not change apart from acquiring a width. The additional absorption $\gammanb$ introduced on the neighboring sites of the defect will destroy the \CT-symmetry of the system itself, but the defect mode is not affected and keeps its structure. 
In contrast varying the eigenfrequency $\nu_d \ne \nu_0$ of the defect resonator will break the \CT-symmetry of the mode leading to non vanishing components $\Psi_n \ne 0$ for $n$ even \cite{mak18}.

Eq.~(\ref{eq:wavefunction}) is depicted in fig.~\ref{fig:eigenstate} as open circles. 
The blue solid line is the experimental eigenfunction obtained by measuring the reflection of the scanning antenna (loop antenna, see fig.~\ref{fig:photo_ant}) above each resonator. 
We can see the staggered structure of the measured wave function. 
Even though the values for even resonators are not zero, still they are small ($|\Psi_n|<10^{-4}|$) close to the defect resonator.
This deviation from the theoretical expression eq.~(\ref{eq:wavefunction}) comes mainly from the variation of the eigenfrequencies of the resonators. 
The increase of the wavefunction at even sites close to the end of the structure is due to finite-size effects and by the fact that the system is coupled to leads.
As the field intensity is small on the neighboring sites of the defect, additional absorption at these resonators will only weakly affect the defect state. 
Thus in the \CT-symmetric case, corresponding to $\nu_d=\nu_0$, the structure will still support a resonant defect mode with high transmittivity.
If we now explicitly violate the \CT-symmetry by modifying the resonant frequency of the defect resonator, eq.~(\ref{eq:wavefunction}) will be perturbed and the wavefunction will acquire weight on the even sites. 
As a result, the resonant defect mode will experience Ohmic losses due to its engagement with the two lossy resonators at the left and right of the defect. 
Consequently the $Q$-factor of the resonant defect mode will deteriorate leading to its destruction via a transition from an under-damping to an overdamping regime. 
Such transition is accompanied by a suppression of the transmittance and an increase to (near-)unity values of the reflectance \cite{mak18}.
The orange dashed line shown in fig.~\ref{fig:eigenstate} corresponds to a frequency shift of the defect resonator of $\Delta\nu=\nu_d-\nu_0=23$\,MHz and an increase of the weight at the neighboring sites of 2 orders of magnitude is visible.

In the next section we will present the experimental results when placing additional absorption on the neighboring sites on the transmittance and the absorption in the limiter.

\section{Experimental results on the \CT-breaking limiter}

\begin{figure}
	\centering 
	\includegraphics[width =\linewidth]{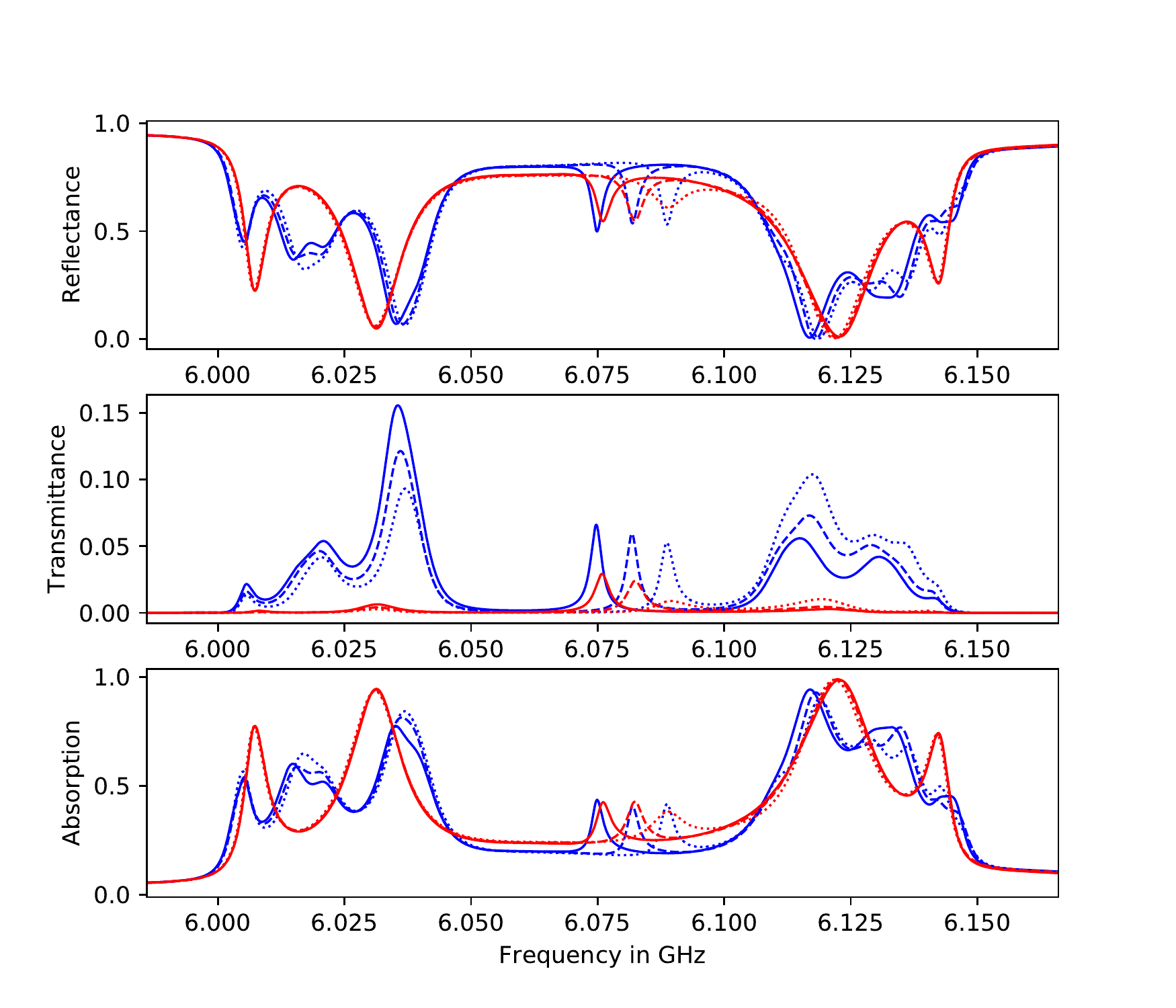}
	\caption{\label{fig:lim_R_T_A}
	Reflectance $R$, transmittance $T$, and absorption $A=1-R-T$ for the 4-D-4 limiter with topological defect at $n$=9 for different frequency shifts of the defect resonators ($\nu_d$) and for different absorption strength $\gammanb$ of the neighboring resonators ($n$=8 and 10).
	The color correspond to different absorption values 
	(blue: $\gammanb = 21$\,MHz; red: $\gammanb = 175$\,MHz) 
	and different frequency-shifts at the central resonator are indicated by the line-style 
	(solid: $\nu_d = 6.069$\,GHz (no shift); dashed: $\nu_d = 6.080$\,GHz; dotted: $\nu_d = 6.092$\,GHz).
	}
\end{figure}

\begin{figure}
	\centering 
	\includegraphics[width =\linewidth]{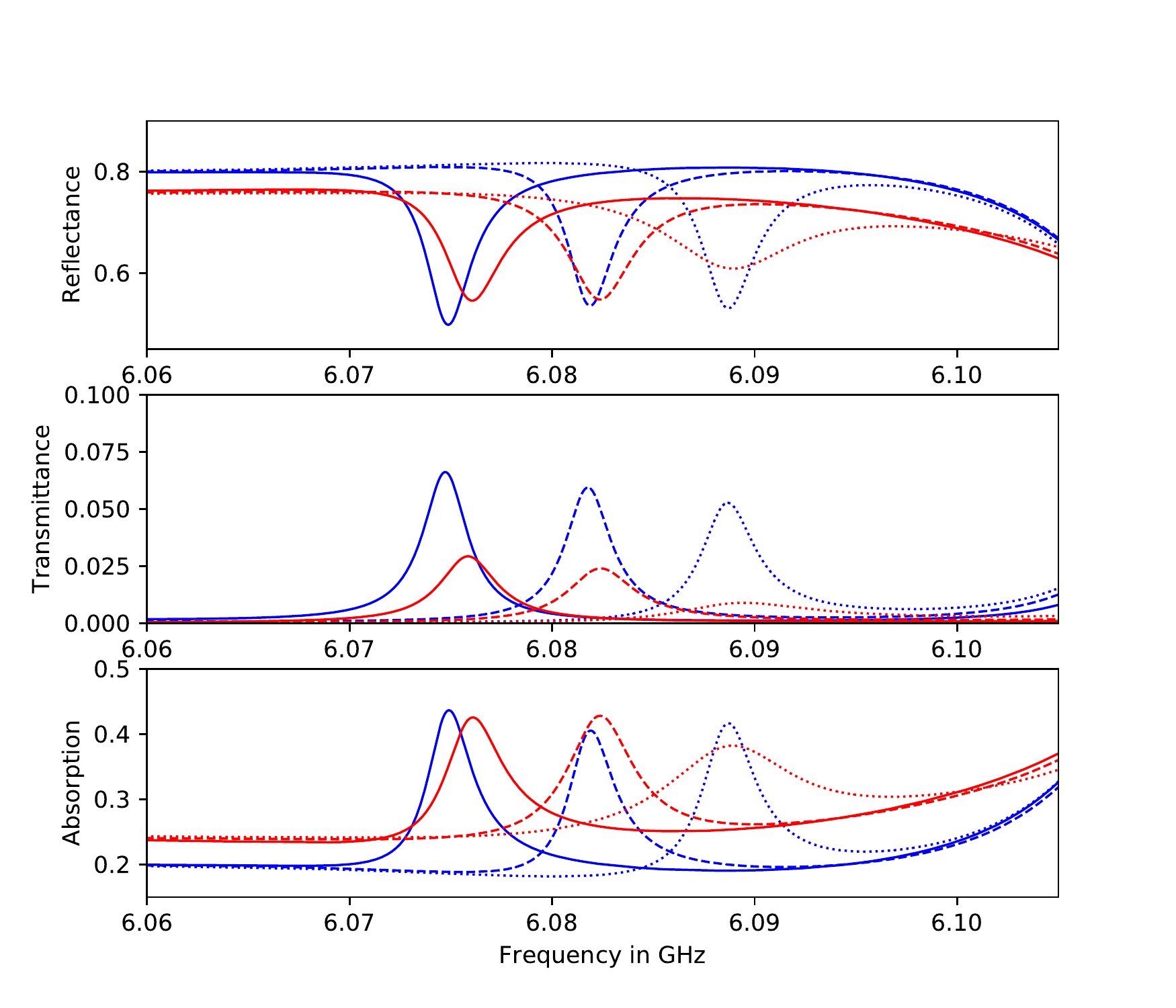}
	\caption{\label{fig:lim_R_T_A_zoom}
		As fig.~\ref{fig:lim_R_T_A} but a zoom to the gap region.
	}
\end{figure}

\begin{figure}
	\centering 
	\includegraphics[width =\linewidth]{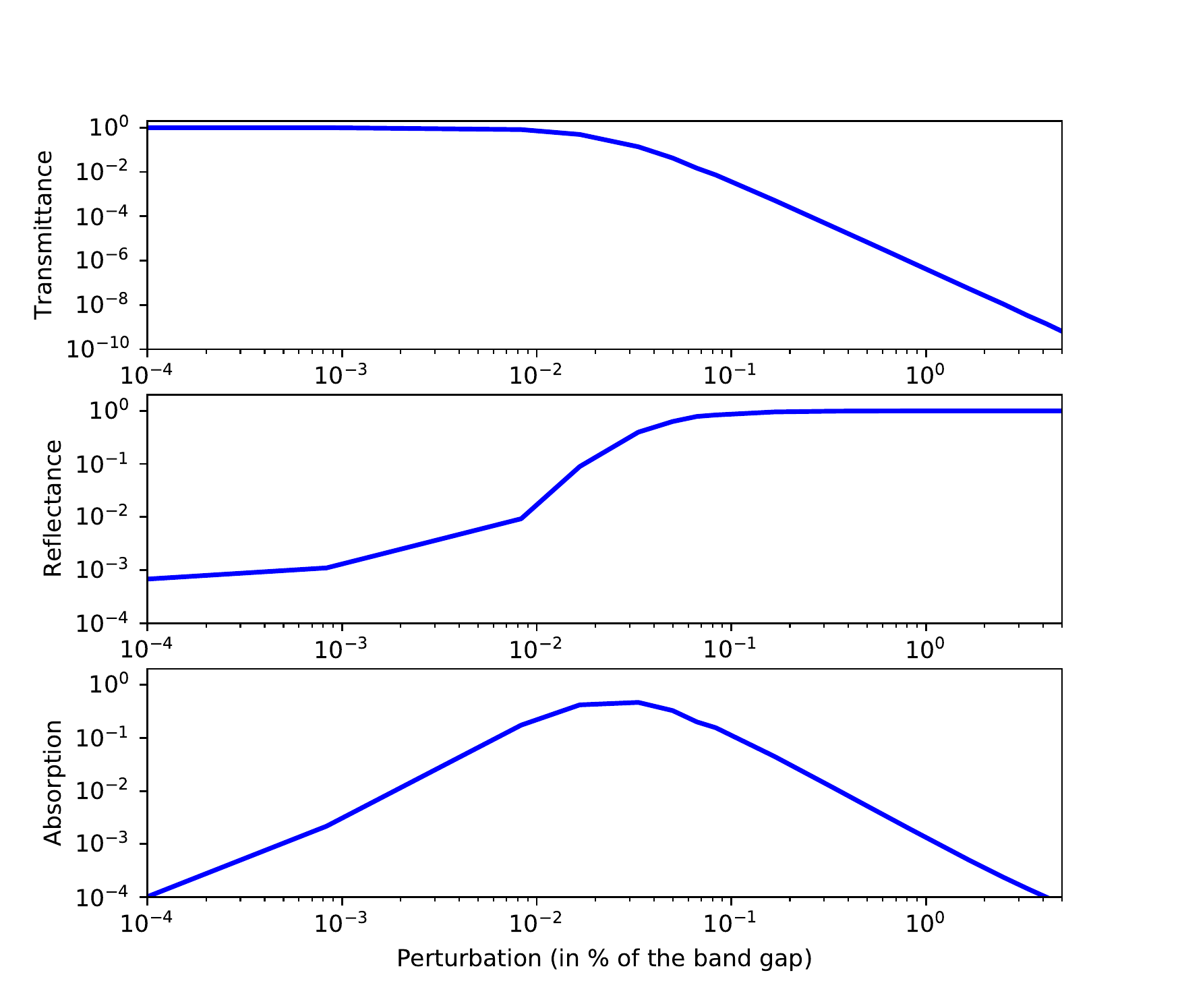}
	\caption{\label{fig:RTA_numerics}
		Transmittance, reflectance and absorption vs. percentile of the perturbation on the defect resonator $n_d=11$ for a $5-D-5$ system.
	}
\end{figure}

Next, we present the experimental results for the transport properties of the \CT-limiter in case that the two lossy resonators are placed on the left and right side of the defect resonator.
For a detailed description of the preparation and characterization of the absorbing resonators see appendix~\ref{sec:absreso}. 
In fig.~\ref{fig:lim_R_T_A} the reflectance $R=|S_{11}|^2$, transmittance $T=|S_{21}|^2$, and absorption $A=1-R-T$ are shown for the 4-D-4 limiter for different frequency shifts of the defect resonators ($\nu_d$) and for different absorption strength $\gammanb$ of the lossy resonators.
For the highest absorption value ($\gammanb = 175$\,MHz, red lines) the background reflectance in the gap is reduced due to the additional strong absorption felt by the resonances within the band. 
To detail the behavior of the resonance we present in fig.~\ref{fig:lim_R_T_A_zoom} only the frequency range of the gap.
In case of the transmittance, we observe the reduction of the resonances with increasing frequency shift [$T_{res}$=0.066, 0.059, 0.053 for $\gammanb=21$\,MHz (blue lines) and 
0.029, 0.024, 0.009 for $\gammanb=175$\,MHz (red lines)], corresponding to a stronger deviation from the \CT-symmetry.
The effect is stronger if the absorption strength is higher but still the transmittance is of the same order of magnitude. 
For the measured reflectance of the limiter the discussion is not as
evident as for the transmittance. 
The resonance heights, widths and frequencies reveal a similar structure in the reflectance as in the transmittance but the baseline in the reflectance for the larger $\gammanb=175$\,MHz is roughly 0.04 smaller thus reducing the minimal reflectance compared to the smaller value
$\gammanb=21$\,MHz. 
Additionally, once approaching the band the baseline is curved, thus leading to a reduction of the increase of the reflectance with increasing $\nu_d$.
As the absorption is mainly displaying the features of the reflectance the main problem resides for it as well.

The experimental findings are significantly less pronounced than predicted by Ref.~\onlinecite{mak18}. 
We attribute this mainly to the losses on the defect resonator and calculated numerically the transmittance, reflectance and absorption of a 5-D-5 limiter with lossless resonators everywhere, but on the neighboring sites of the defect, where we assumed losses of $\gammanb=500$\,MHz.
The couplings used are $t_1=50$\,MHz and $t_2=10$\,MHz, thus the band-gap is broader in this case.
For the calculation the scattering matrix formalism is used, mimicking each antenna coupled to the 5-D-5 limiter by a one-dimensional semi-infinite tight-binding lattice (for the efficiency of this modeling see \cite{mak18}).

In fig.~\ref{fig:RTA_numerics},	we plot maximal transmittance $T$,	minimal reflectance $R$, and maximal absorption $A$ within the band-gap, as a function of the relative changes of $\nu_d$. 
We observe a drop of more than 10 orders of magnitude in transmittance and about 4 orders of magnitude in absorption for large perturbation values equal to a few percentile points of the band-gap, much smaller than the ones implemented in the experiment. 
Thus we conclude the problem in the performance of the experimental set-up is associated with the moderate $Q$-factor and associated losses of the resonators used in the experiment and predominantly of the defect resonator where the losses are amplified due to the exponentially large values of the field at $n=n_d$ [see eq.~(\ref{eq:wavefunction})].

\section*{Acknowledgments}
T.K. acknowledge partial support from the Office of Naval Research via grants N00014-16-1-2803 and N00014-19-1-2480 and from DARPA NLM program via grant No. HR00111820042.

\appendix

\section{Single resonator and experimental precision}
\label{sec:singlereso}

Here we present details on the single resonator and the experimental precision of the measurements.

To derive an analytic expression for the electric and magnetic field we assume a dielectric cylinder with an index of refraction $n$ and height $h$ sandwiched between two perfectly conducting metallic plates both touching the resonator. 
Then the first transverse electric resonance TE$_1$, i.e. magnetic field pointing in the $z$ direction, reads:
\begin{equation}
	\vec{B}=B_z(x,y,z) e^{i\omega t} \evec_z = \varphi_B \left( x, y \right) \sin\left(\frac{ \pi}{h} z\right) e^{i\omega t} \evec_z 
\end{equation}
$\varphi_B$ can be deduced from the 2-dimensional Helmholtz equation in polar-coordinates 
\begin{equation}
\left( \Delta_{\text{polar}} + k_\perp^2 \right) \varphi_B \left(r,\theta \right) = 0
\end{equation}
with $k_{\perp} = \sqrt{n \omega^2/c^2 - k_z^2} = \sqrt{k^2 - k_z^2}$, where $k_z = \frac{\pi}{h}$ the corresponding wavenumber in $z$-direction. 
For the used height of $h = 12$\,mm, an index of refraction of $n \approx 6$ for the resonators and the used frequency around 6\,GHz, the wave number $k_{\perp}$ inside the resonator ($r<r_D$) is real, but outside ($r > r_D$) it is purely imaginary, leading to evanescent fields, and therefore to an evanescent coupling between the resonators. Taking into account the continuity-condition for the field at the resonator-air interface we find 
\begin{equation}
\vec{B} \left(r, z \right)= \left(0,0,B_z \right) = \begin{cases}
B_0 \sin \frac{\pi z}{h} J_0(k_\perp r) \evec_z\text{ for } r< r_D \\
\alpha B_0 \sin \frac{\pi z}{h} K_0(\gamma r) \evec_z \text{ for } r> r_D \\
\end{cases}
\label{eqn_Bfield_ideal_singleDisk}
\end{equation}
where $r$ is the distance from the center of the resonator, $r_D$ the radius of the resonator, ${k_\perp = \sqrt{\left(\frac{2 \pi \nu_0 n}{c_0}\right)^2 - \left(\frac{\pi}{h} \right)^2}}$ and $\gamma= \sqrt{\left(\frac{\pi}{h} \right)^2 -\left(\frac{2 \pi \nu_0}{c_0}\right)^2}$. $J_0$ and $K_0$ are Bessel functions and $B_0$ and $\alpha$ are constants, where $\alpha$ is to be determined by the continuity equation at the surface of the resonator.

The corresponding electrical field $\vec{E}\left(\vec{r}, t \right) = \vec{E}\left( \vec{r} \right) \text{e}^{\text{i} \omega t}$ with harmonic time dependency is thus given by
\begin{equation}
\vec{E} \left(r,z \right) \approx 
\begin{cases}
E_0 \sin \frac{\pi z}{h} J_1(k_\perp r) \evec_\theta &\text{ for } r< r_D \\
\alpha E_0 \sin \frac{\pi z}{h} K_1(\gamma r) \evec_\theta &\text{ for } r> r_D \\
\end{cases}
\label{eqn_Efield_ideal_singleDisk}
\end{equation}
with $E_0 = - \frac{i \mu_0 B_0 c_0}{2 \pi \nu_0}$. 

\begin{figure}
	\centering 
	\includegraphics[width =\linewidth]{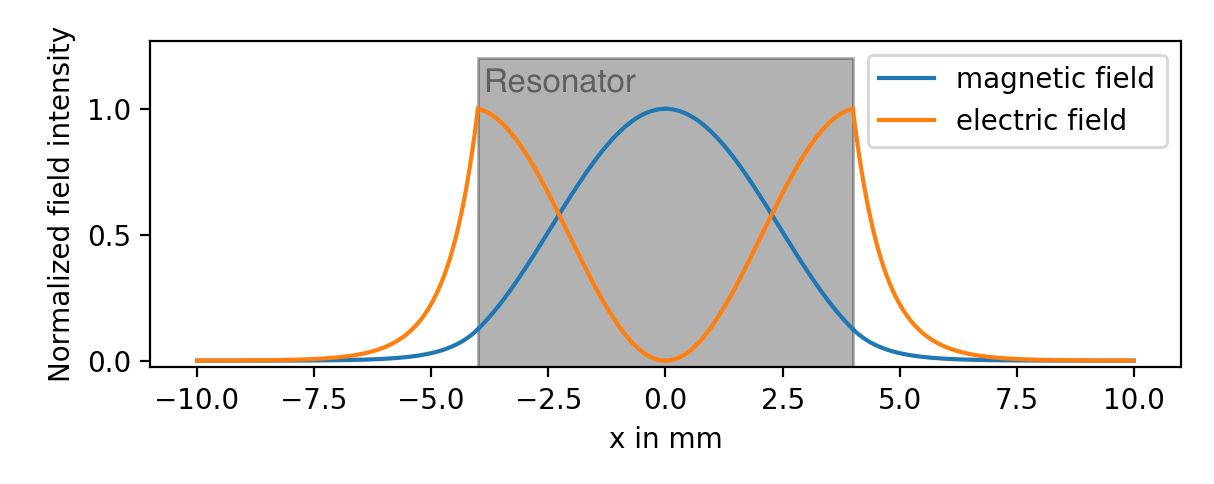}
	\caption{\label{fig:single_res_fields}
		The electric (red solid line) and magnetic (blue solid line) field are sketched for a dielectric cylindrical resonator (shaded area). 
	}
\end{figure}

The derived field intensities for the electric and magnetic field intensities are plotted in fig.~\ref{fig:single_res_fields}. 
One can see that the magnetic field has its maximum at the center of the resonator, whereas the electric field is maximal at the edge of the resonator. 
The energy is well localized inside of the resonator while outside the fields are decreasing fast with increasing distance from the center. 
This is an important property in order to be able to describe the system with the tight-binding formalism.
Even though the experimental setup has a gap between the top plate and the resonator the essential description is still valid mainly modifying the existence of several evanescent modes outside. 
For details see Ref.~\onlinecite{bel13a}. 

\begin{figure}
	\centering 
	\includegraphics[width =\linewidth]{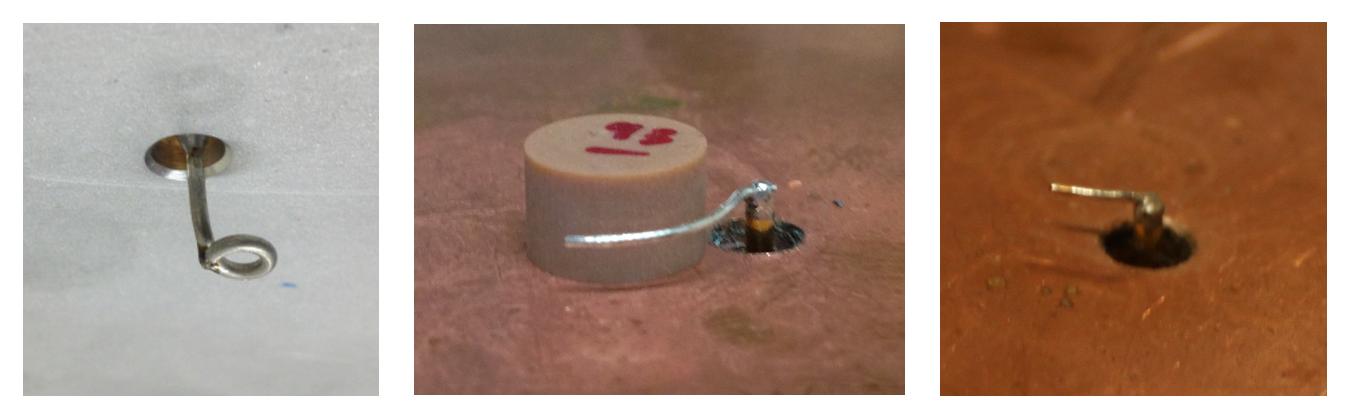}
	\caption{\label{fig:photo_ant}
		(left) Loop antenna positioned on top of the resonator to couple to the magnetic field.
		(center) Curled kinked antenna to guarantee strong coupling to the electric field.
		(right) Small kink antenna gives rise to weak coupling to the electric field of the resonator.		
	}
\end{figure}

From the structure of the field the choice of different antennas becomes evident. 
In fig.~\ref{fig:photo_ant} different types of antennas, which have been used, are shown. 
The curled kinked excitation antenna situated roughly at half the height of the resonators and curling around half the resonator is a good choice to give rise to strong coupling. 
Additionaly, the scanning antenna, a loop antenna coupling to the magnetic field, is optimal to get information of $B_0$ which is sufficient to obtain the field in the $xy$-plane. 

\begin{figure}
	\centering 
	\includegraphics[width =\linewidth]{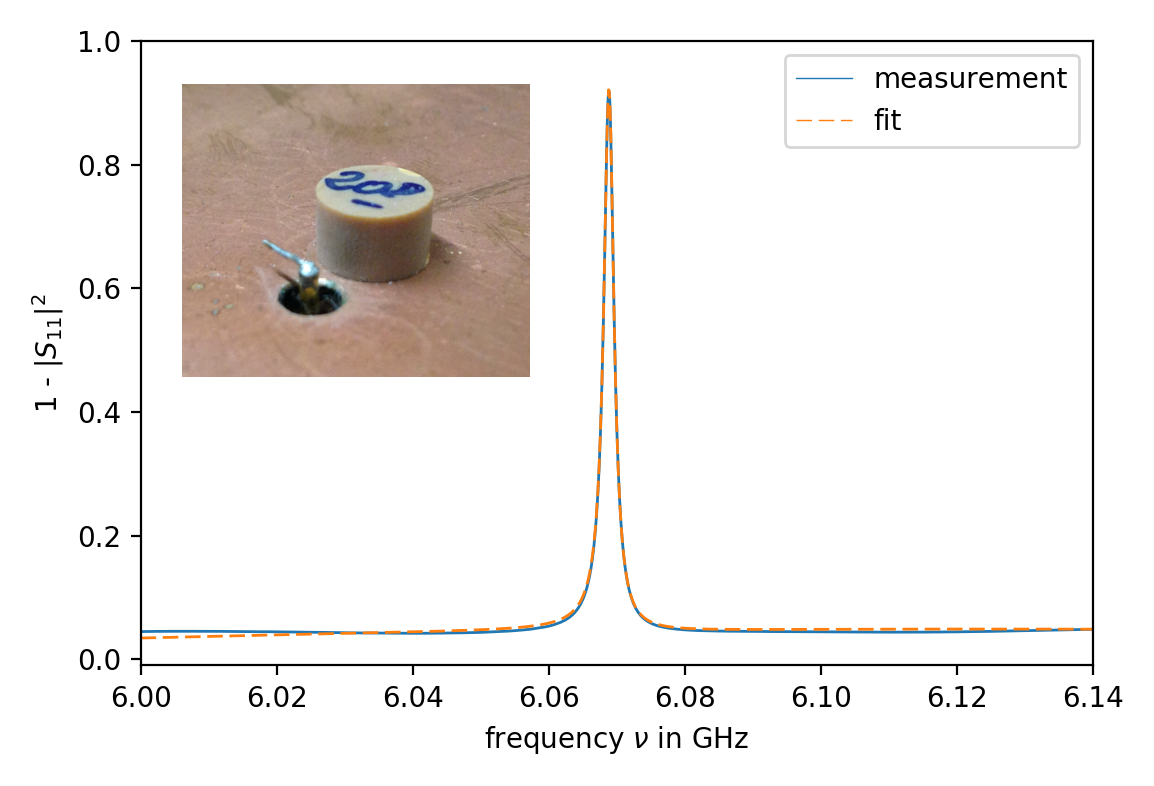}
	\caption{\label{fig:single_resonance}
		Measured reflectance (blue solid line) showing the TE$_1$ resonance of the dielectric resonator at around 6.069\,GHz. 
		The dashed line corresponds to a Lorentzian fit including a linear background [see eq.~(\ref{eq:fitfunc})]. 
		The inset show the experimental setup using a weakly coupled kink antenna for this measurement.
	}
\end{figure}

Figure~\ref{fig:single_resonance} shows the reflectance $|S_{11}|^2$ as a function of frequency. A resonance is found around 6.069\,GHz which is perfectly described by a Lorentzian. 
The fit is taking into account the background by via a complex linear function
\begin{equation}\label{eq:fitfunc}
  S_{11}=1 - \frac{a}{\nu-(\nu_0-i\gamma/2)}+ c_1 \nu + c_2
\end{equation}
where $\nu_0$ is the eigenfrequency of the resonator and $\gamma$ its width. $c_1$ and $c_2$ are complex constants. 
The fit shown in fig.~\ref{fig:single_resonance} gives $\nu_0=6.069$\,GHz and $\gamma=1.8$\,MHz. 
To generate a dimer chain we need that the eigenfrequencies of the resonators are sufficiently close, i.e., within $\gamma$. 
To verify this we performed several test measurements.

\begin{figure}
	\centering 
	\mbox{
		\includegraphics[height=2cm]{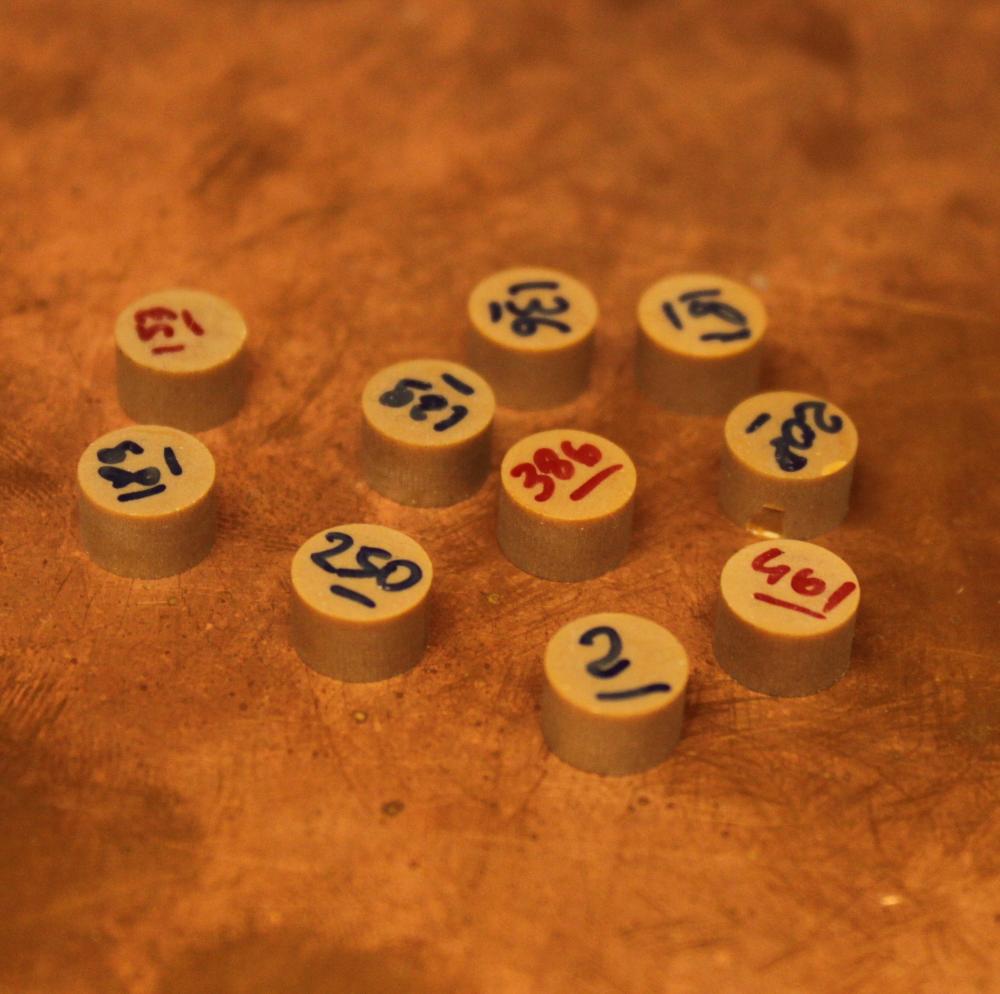}
		\includegraphics[height=2cm]{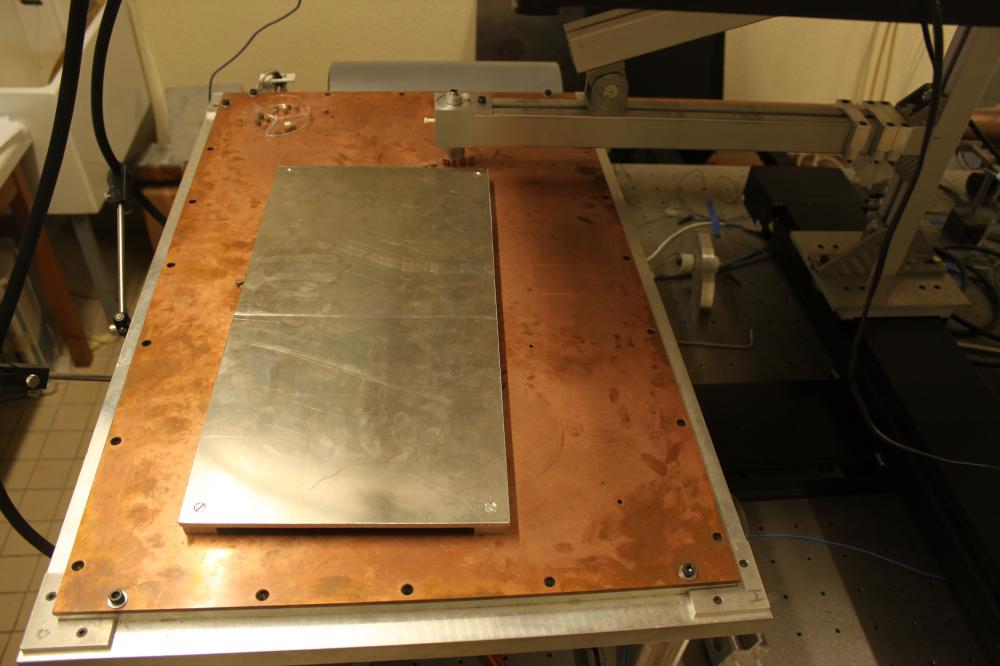}
		\includegraphics[height=2cm]{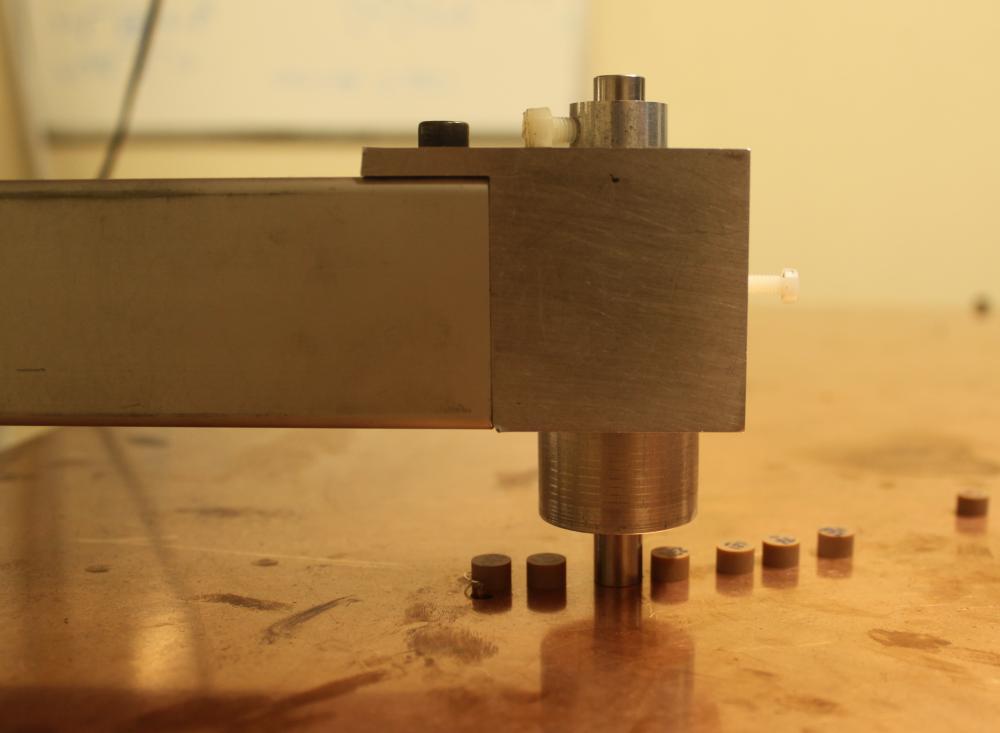}
	}
	\caption{\label{fig:photo_plate}
		(left) Photograph of the labelled resonators.
		(center) Aluminum top plate placed over a single resonator with a distance of 12\,mm.
		(right) Liftable downpipe at the end of the movable arm used to precisely place the resonators.
			}
\end{figure}

\begin{figure}
	\centering 
	\includegraphics[width =.99\linewidth]{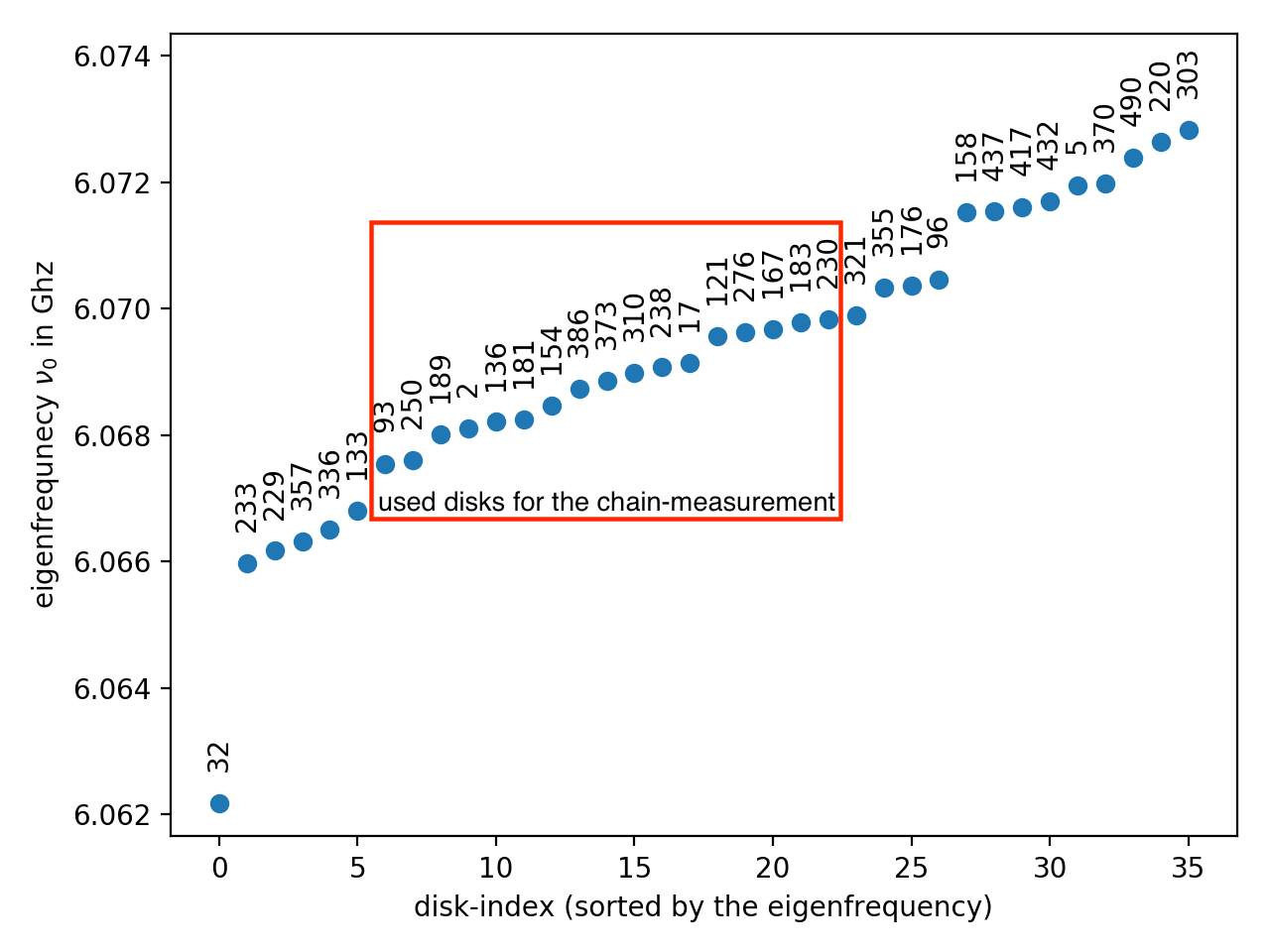}\\
	\includegraphics[width =.99\linewidth]{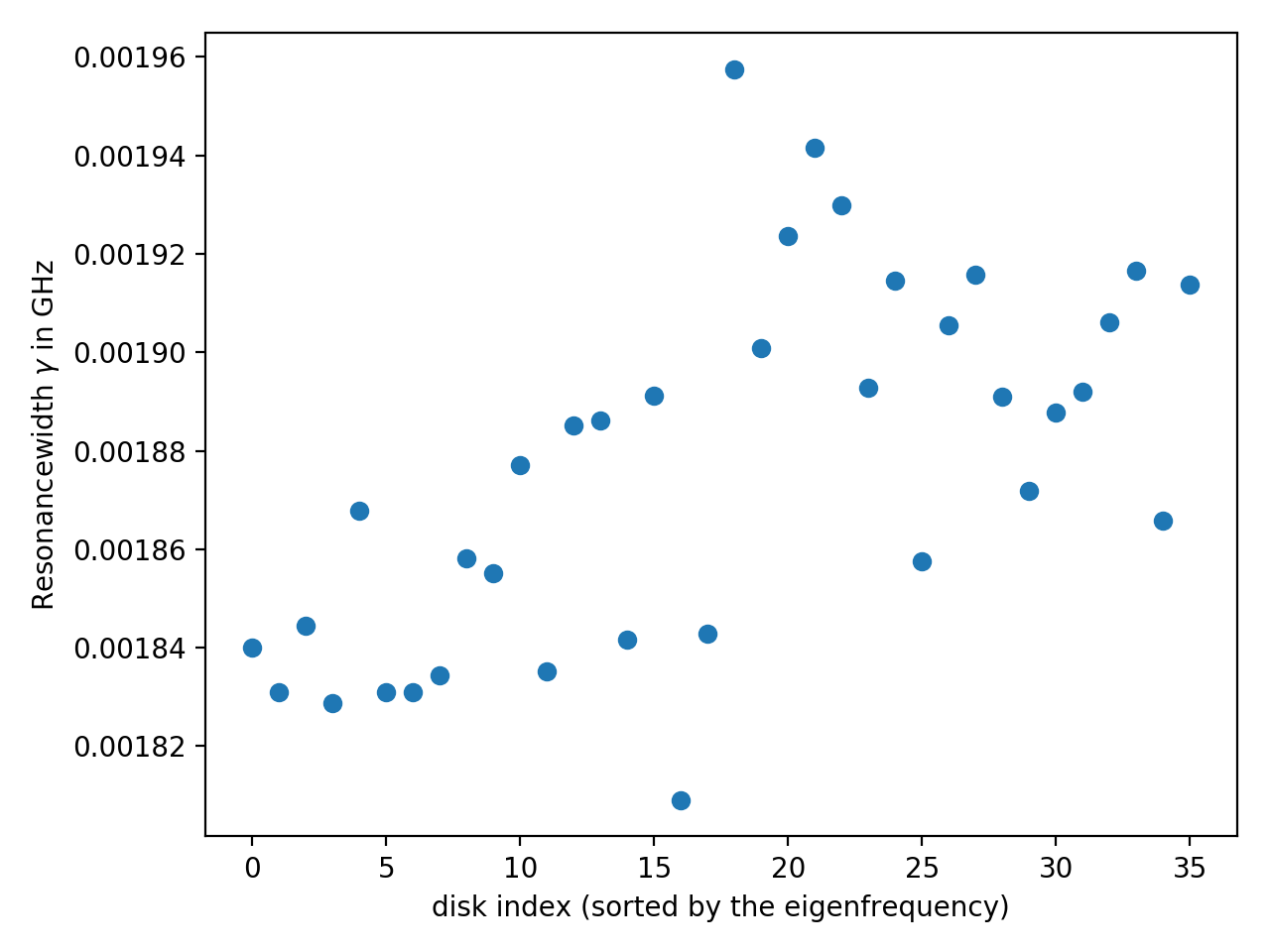}
	\caption{\label{fig:nu_0_gamma_dependence}
		Eigenfrequency and width of the resonances measured for each resonator.
	}
\end{figure}

First of all, we labeled all resonators [see fig.~\ref{fig:photo_plate}(left)] and measured the reflection $S_{11}$ and fitted them by a complex Lorentzian. 
In fig.~\ref{fig:nu_0_gamma_dependence} we present the found resonance frequencies and widths. 
We choose the resonators which lie within a 2\,MHz wide frequency window corresponding to the resonance width found of about 1.8\,MHz. 

\begin{figure}
	\centering 
	\includegraphics[height=.29\linewidth]{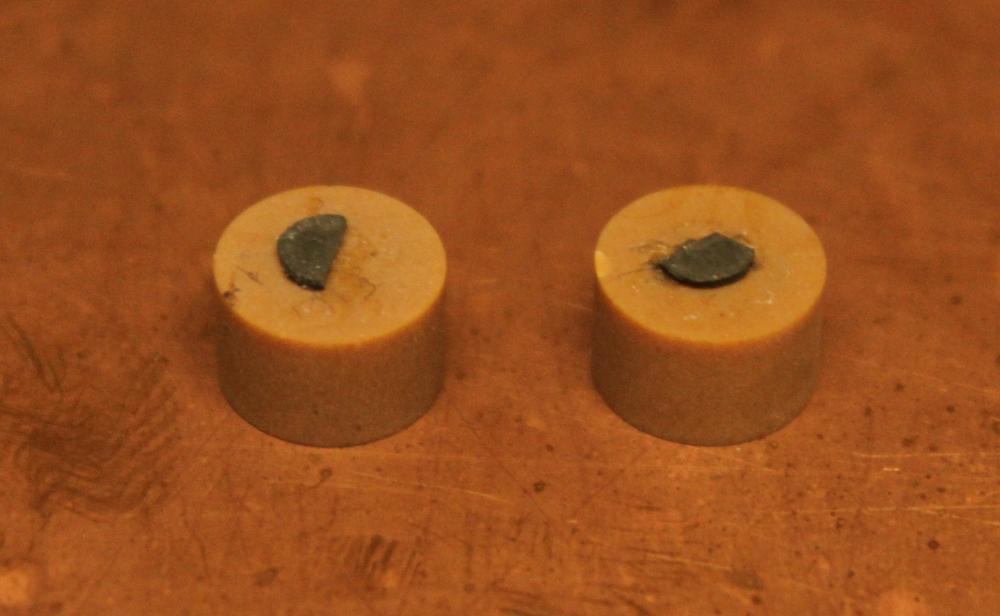}
	\includegraphics[height=.29\linewidth]{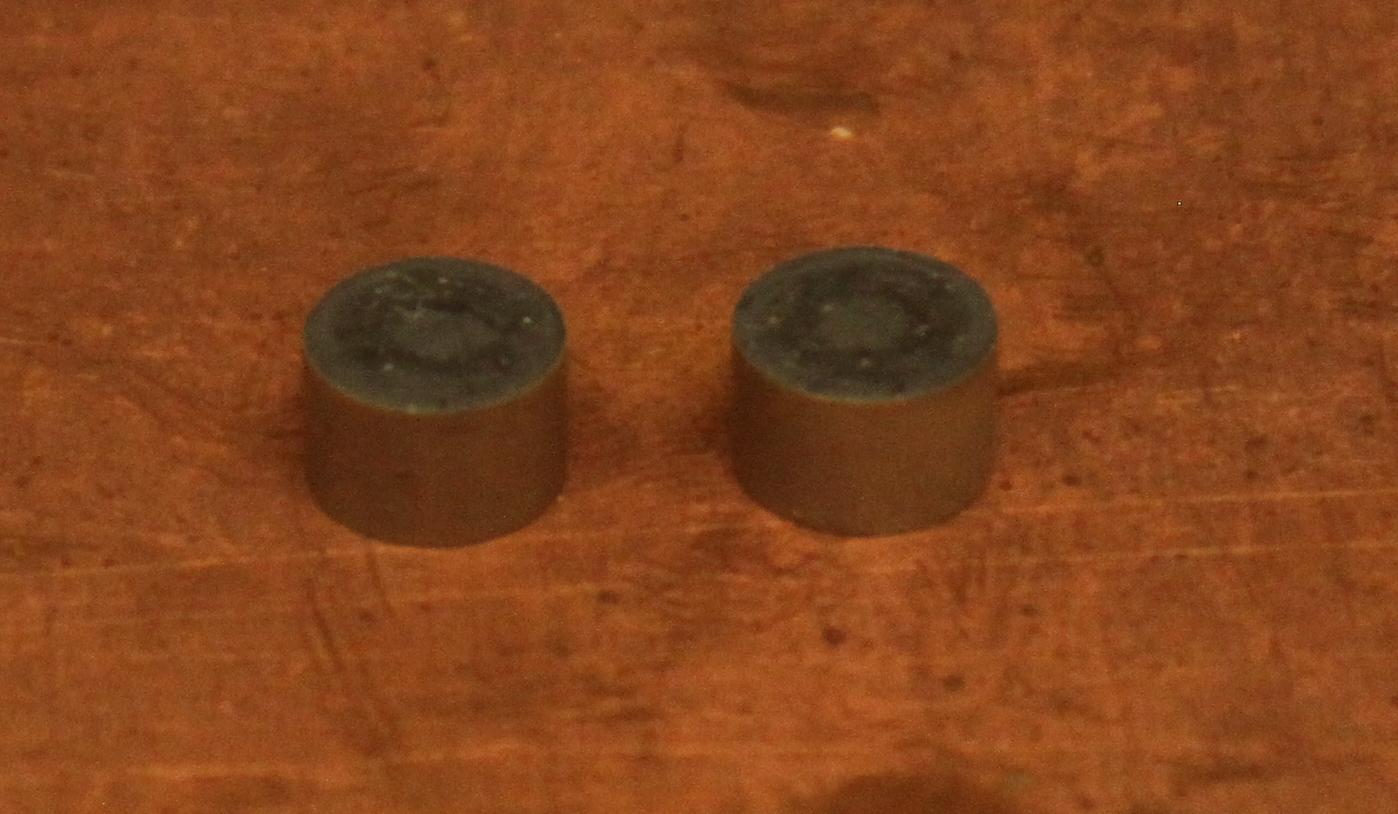}
	\caption{\label{fig:photo_abs}
		(left) Photograph of the two resonators with elastomer absorber patches glued on top.
		(right) Photograph of the two resonators with a conductive graphite layer sprayed on top.
	}
\end{figure}
\begin{figure}
	\centering 
	\includegraphics[width =\linewidth]{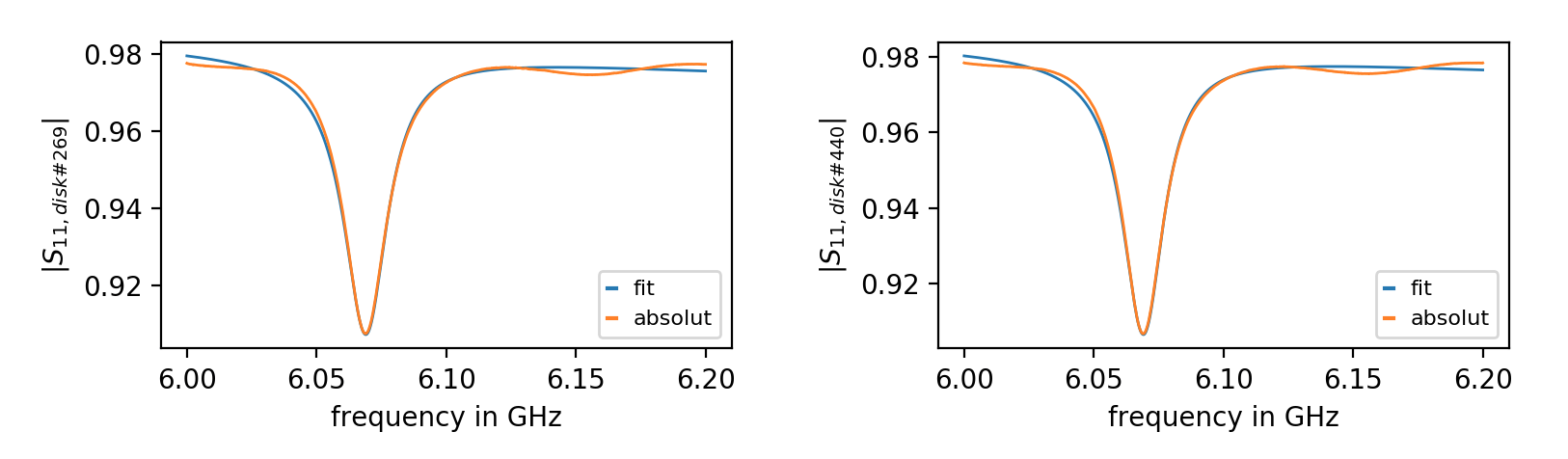}
	\includegraphics[width =\linewidth]{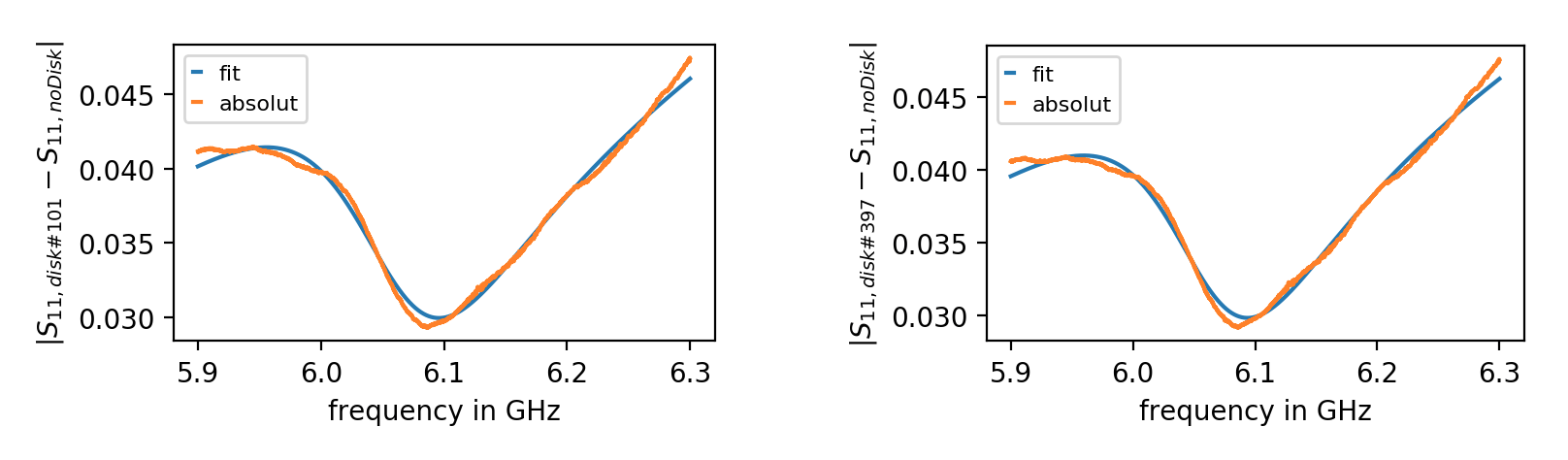}
	\caption{\label{fig:resonance_abs}
		Reflectance measurements of the absorbing resonators shown in fig.~\ref{fig:photo_abs}. 
		(Top) Resonance for the two resonators with elastomer patches.
		(Bottom) Resonance for the two resonators with graphite layer.
	}
\end{figure}

Next we repeated measurements where we lift the top plate [see fig.~\ref{fig:photo_plate}(center)] and place it again at the same position above the resonator. This gave rise to a variation of 0.05\,MHz of the resonance far below the fluctuation already present. 
Now we placed the same resonator several times at the same position using a lift-able down-pipe at the end of the movable arm [see fig.~\ref{fig:photo_plate}(right)]. 
The arm is controlled by a high precision $xy$-stage leading to a position error of less than 0.1\,mm.
The resonators are additionally pushed down using a stick with a rubber patch at its end. 
The variation is now of the order of 0.35\,MHz but still less than the resonance width. 
Overall the variation of the eigenfrequencies due to the chosen resonator, its placement, and the positioning of the top plate is sufficiently small to keep the topological structure necessary for the operation of the limiter intact.

As one can see from fig.~\ref{fig:nu_0_gamma_dependence} it is easily possible to perform a perturbation of the eigenfrequency of the defect resonator by an appropriate choice of the resonator. 
For the perturbed resonators at the defect site we choose later resonator no.~497 with an eigenfrequency of $\nu_0 = 6.092$\,GHz and no.~10 with an eigenfrequency of $\nu_0 = 6.080$\,GHz both not inside the frequency window presented in fig.~\ref{fig:single_resonance}. 

Furthermore we need two absorbing resonators next to the topological defect. 
This will be described in the following section.

\section{Absorbing resonators}
\label{sec:absreso}

The limiter is based on breaking the \CT-symmetry by shifting the resonance frequency of the defect resonator. 
To finally reduce the transmittance this frequency shift needs to be transformed into an absorption which gives finally rise to the enhanced reflectance. 
This absorption is induced by adding absorbing material to the resonators neighboring the topological defect. 
On the one hand side the resonance frequency of these resonators need to stay at the same frequency but they should show a strongly enhanced width. 

Adding the absorbing material will change the eigenfrequency of the resonator. This needs to be compensated by choosing a resonator which has the corresponding eigenfrequency. 
We use here two different materials to enhance the absorption, either elastomer patches, which reduces the eigenfrequency, or graphite spray, which increases the eigenfrequency, (see fig.~\ref{fig:photo_abs}). 
In fig.~\ref{fig:resonance_abs} reflectance measurements for these four resonators are shown. 

To perform a fit for the two graphite layer resonators we had to subtract the reflections measurements when no resonator was present. 
For all four resonators the eigenfrequencies are close to the 6.069\,GHz of the resonators without absorption compared to their width.
Details on the resonance frequencies and width are shown in tab.~\ref{tab:resfrq}. 
By means of the absorbing material we can rise the resonance width by up-to a factor of 100.

\begin{table}[h]
	\begin{tabular}{|r|c|c|c|}\hline
		No. & Absorber & $\nu_0$/GHz & $\gamma$/MHz\\\hline
		373 & no        & 6.069 &  1.8\\\hline
		269 & elastomer & 6.069 &  21\\\hline
		440 & elastomer & 6.069 &  20\\\hline
		101 & graphite  & 6.054 & 180\\\hline
		397 & graphite  & 6.056 & 170\\\hline
		10  & no        & 6.080 &  1.8\\\hline
		497 & no        & 6.092 &  1.8\\\hline
	\end{tabular}
	\caption{\label{tab:resfrq}
		Resonance frequencies and width for the different kind of resonators used.}
\end{table}

\end{document}